\documentclass[prl,reprint]{revtex4-1}

\usepackage{graphicx, epsfig, amssymb, amsmath, lineno, subfigure, float,lmodern}
\newcommand{\mb}{\mathbf}

\begin{document}
\pagenumbering{roman}
\title{Eddy current interactions in a Ferromagnet-Normal metal bilayer structure, and its impact on ferromagnetic resonance lineshapes  }

\author{Vegard Flovik}
\email{vegard.flovik@ntnu.no}
\affiliation{Department of Physics, Norwegian University of Science and
  Technology, N-7491 Trondheim, Norway}

\author{Ferran Maci\`{a}}
\affiliation{Grup de Magnetisme, Dept. de F\'isica Fonamental, Universitat de Barcelona, Spain}

\author{Andrew D. Kent}
\affiliation{Department of Physics, New York University, 4 Washington Place, New York, New York 10003, USA }

\author{Erik Wahlstr\"{o}m}
\affiliation{Department of Physics, Norwegian University of Science and
  Technology, N-7491 Trondheim, Norway}

\date{\today}

\begin{abstract}

We investigate the effect of eddy currents on ferromagnetic resonance (FMR) in ferromagnet-normal metal (FM/NM) bilayer structures. Eddy-current effects are usually neglected for NM layer thicknesses below the microwave (MW) skin depth ($\simeq 800$ nm for Au at 10 GHz). However, we show that in much thinner NM layers (10-100 nm of Au or Cu) they induce a phase shift in the FMR excitation when the MW driving field has a component perpendicular to the sample plane. This results in a strong asymmetry of the measured absorption lines. In contrast to typical eddy-current effects, the asymmetry is larger for thinner NM layers and is tunable through changing the sample geometry and the NM layer thickness.

\end{abstract}

\pacs{}

\maketitle

\section{Introduction}
Eddy currents are induced currents in conductors by changing magnetic fields. These currents flow in closed loops perpendicular to the driving fields, and produce additional Oersted fields that partially compensate the external driving fields. The effects of eddy currents on FMR in conducting films is well known in the limit of film thickness approaching their electro-magnetic skin depth ( $\simeq$ 800 nm for bulk Au at 10 GHz). In those cases, eddy-current effects can lead to linewidth broadening and give rise to spin-wave excitations due to inhomogenous microwave fields \cite{eddy1,eddy2}. 

The microwave frequency spin dynamics in nanostructures usually involves stacks of layers combining FM and NM at the nanometer scale \cite{spindyn1,spindyn2}. Although eddy-current effects are usually neglected in metals with thicknesses below their skin depth, some studies have shown that this may be important also for normal metal (NM) films far below their skin depth \cite{screening,screening2,screening_FMR, screening_FMR2, screening_FMR3, screening_FMR4}. In these studies it was predominantly microwave-screening effects that were considered, and little attention was paid to how the induced Oersted fields can affect the magnetization dynamics in an adjacent ferromagnetic thin film.

Ferromagnetic resonance (FMR) spectroscopy experiments probe static and dynamic properties of magnetic materials. The technique relies on measuring the microwave absorption associated to the precession of the magnetization. In FMR experiments, position and width of absorption lines carry valuable information about material parameters such as anisotropy fields and magnetic damping \cite{review_FMR}.

Many experimental setups used for such studies have the main component of the MW driving field oriented in the sample plane (coplanar waveguide (CPW)/stripline FMR). Non-uniformity of the MW field, and sample position with respect to the CPW/stripline center would still lead to field components perpendicular to the sample plane, which would enhance the effects of eddy currents. 
On the other hand, in cavity-FMR setups, the MW fields can be oriented either parallel or perpendicular to the sample plane depending on the cavity.

Differences in symmetry of FMR lines have been used to study the spin pumping from a magnetic material to a normal metal \cite{spinHall,spinHall2,spinHall3}. We notice here that a recent study has reported different values for the voltage induced by the inverse spin hall effect, depending on the cavity mode used \cite{spinHall4}. In such studies, lineshape symmetry is one of the main parameters used to analyze the results. 

Hence, to correctly interpret experimental data involving FMR it is important to understand how eddy currents---even in very thin films---can cause modifications in the measured FMR lineshape. 

In this study we investigate the contribution of eddy currents to the FMR absorption lineshapes in ferromagnet-normal metal (FM/NM) bilayer structures. We have systematically studied how the sample geometry and NM thickness affects the coupling between microwave (MW) fields and eddy-current-induced fields, and we show that this coupling is tunable through changing the sample geometry and the NM layer thickness.

\section{Theoretical model for the observed lineshapes}

\begin{figure}[ht]
\centering
\includegraphics[width=80mm]{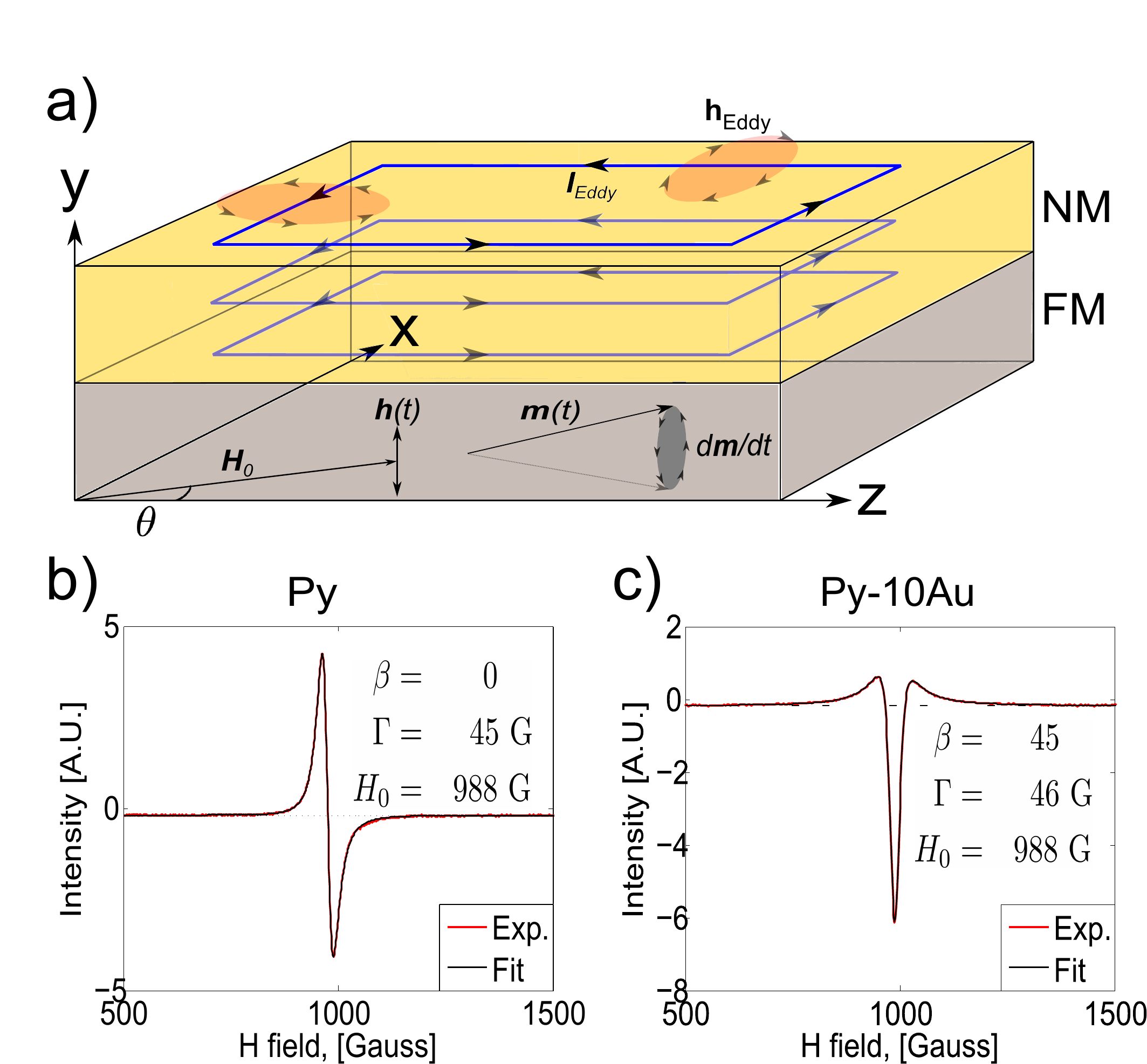}
\caption{\footnotesize (a) A schematic of the sample geometry showing the path of the induced eddy currents flowing in closed loops around the sample, with highest current density along the sample edges \cite{ind_eddycurr, ind_eddycurr2}. (b) and (c):  FMR lineshapes and fitted parameters from Eq.(\ref{eq:fiteq2}) for a sample with 10 nm Py (b), and the same sample after being capped with 10 nm Au (c).}
\label{fig:lineshapes_sampgeom}
\end{figure}

The ferromagnetic resonance is usually driven directly by the MW field from a cavity or from a coplanar waveguide/microstrip line. However, capping a FM sample with a NM layer leads to circulating eddy currents in the NM, and additional Oersted fields in the FM. These Oersted fields have a different phase with respect to the MW fields---there is a relative phase lag between the MW fields and the Oersted fields from the induced currents---, and this results in a distortion of the FMR lineshape. A sketch of the FM/NM bilayer geometry and the path of the induced eddy currents is shown in Fig. \ref{fig:lineshapes_sampgeom}a. The induced currents flow in closed loops in the sample plane, with highest current density along the sample edges \cite{ind_eddycurr,ind_eddycurr2}. Figures \ref{fig:lineshapes_sampgeom}b and c compares two representative FMR lineshapes for a 10 nm Py sample before and after capping it with 10 nm Au; although resonance frequency and linewidth stay constant, the lineshape changes considerably.

To understand the origin of the distorted lineshapes due to the induced eddy currents, we consider a model describing the magnetization dynamics of the FM, starting from the Landau Lifshitz Gilbert equation \cite{LLG}:

\begin{equation}\label{eq:LLG}
\frac{\partial \mb{M}}{\partial t} = -\gamma \mb{M} \times \mb{H}_{\text{eff}} + \frac{\alpha}{M_s}\Bigg ( \mb{M} \times \frac{\partial \mb{M}}{\partial t } \Bigg),
\end{equation}

\noindent
where $\gamma$ is the gyromagnetic ratio, $\alpha$ is the Gilbert damping parameter, and $M_s$ is the saturation magnetization. The effective magnetic field, $\mb{H}_{\text{eff}}$, includes the external field $\mb{H}_0$, the anisotropy field, $\mb{H}_A$ (we neglect the effects of dipole and exchange fields) and a driving oscillatory field, $\mb{h}_{\text{ac}}$, composed of MW fields and fields from the eddy currents. In the following, the contribution from the anisotropy field to the effective field is neglected, as this is negligible compared to the resonance field for Py. 

We assume the external applied field, $\mb{H}_0$, is along $\mb{z}$ in the film plane (see Fig. \ref{fig:lineshapes_sampgeom}a) and only consider perturbations of the oscillatory field, $\mb{h}_{\text{ac}}$, in the x-y plane, perpendicular to the external applied field (the components in the direction of the applied field do not directly perturb the dynamics of $\mb{M}$). 

The phase of the microwave excitation in any FMR experiment is arbitrary and can depend on many factors. However, as we are only interested in relative phase differences, we can set the reference phase of the MW field to zero. The combined driving field can thus be written in the form:

\begin{equation}\label{eq:beta}
\begin{split}
 \mb{h}_{\text{ac}}(t)=\mb{h}_{\text{MW}} e^{i \omega t} + \mb{h}_{\text{ind}} e^{i (\omega t - \phi)}  \\
= [\mb{h}_{\text{MW}} +  \mb{h}_{\text{ind}} (\cos \phi - i \sin \phi) ]e^{i \omega t } 
 \equiv \mb{h} e^{i \omega t }[1 -  \mb{\beta} i],
\end{split}
\end{equation}

\noindent
where $\phi$ is the relative phase difference between the MW field and the induced field,  and $\mb{h}= \mb{h}_{\text{MW}}+  \mb{h}_{\text{ind}} \cos \phi$. The parameter $\beta$ is thus defined as:

\begin{equation}\label{eq:beta2}
\beta=  ( \mb{h}_{\text{ind}} \sin \phi) / ( \mb{h}_{\text{MW}} + \mb{h}_{\text{ind}} \cos \phi )=(\beta_x , \beta_y),
\end{equation}

\noindent 
and accounts for the relative magnitude of the two fields and their phases in the $\mb{x}/\mb{y}$ direction respectively. The parameter $\beta$ will thus approach zero when the induced field is small compared to the MW field, or the phase difference between the MW field and the induced field is close to 0 degrees. There will also be a maxima for $\beta$ for some value of the phase difference in the range between 90-180 degrees, which depends on the magnitude of the induced field compared to the MW field.  

The magnetization $\mb{M}(t)$ is then taken of the form $\mb{M}(t)=M_0 \mb{z} + \mb{m} e^{i\omega t}$, where $\mb{m} \perp \mb{z}$. The magnetic response to small excitation fields, $\mb{m}=\boldsymbol \chi \mb{h}$, is determined by the Polder susceptibility tensor $\boldsymbol \chi$ \cite{polder}.

The elements of $\boldsymbol \chi$ were determined by solving Eq. (\ref{eq:LLG}), and discarding higher order terms.
Setting $\mb{m}=\boldsymbol \chi \mb{h}$ and introducing $\omega_0=\gamma H$ and $\omega_M=\gamma M_0$, one obtains:

\begin{equation}\label{polder}
\boldsymbol \chi=
\begin{pmatrix}
\chi_{xx} & i \chi_{xy} \\
-i \chi_{yx} & \chi_{yy}  \\
\end{pmatrix},
\end{equation} 

\noindent 
where the matrix elements are given by

\begin{equation}\label{eq:polder2}
\chi_{xx /yy}= \frac{(1- i \beta_{x/y})\omega_M (\omega_0 + i \alpha \omega)}{\omega_0^2 - \omega^2 (1+\alpha^2) + 2 i \alpha \omega \omega_0},
\end{equation}

\begin{equation}\label{eq:polder3}
\chi_{xy /yx}= \frac{(1- i \beta_{y/x} )\omega \omega_M}{\omega_0^2 - \omega^2 (1+\alpha^2) + 2 i \alpha \omega \omega_0}.
\end{equation}

The observable quantity in our FMR experiments is the MW power absorption, which is given by an integral over the sample volume $V$ \cite{jackson}:

\begin{equation}\label{eq:Pabs}
P_\text{abs} = \frac{1}{2} \Re \int_V  i \omega (\boldsymbol \chi \mb{h})\cdot \mb{h}^*  dV
\end{equation}

Splitting $\boldsymbol \chi$ into its real and imaginary part and using that $h_x$ and $h_y$ are orthogonal, one obtains:

\begin{equation}\label{eq:Pabs2}
\begin{split}
P_\text{abs}= \frac{1}{2} \Re \int_V  i \omega [\boldsymbol \chi^{'} + i \boldsymbol \chi^{''}]
\left(
\begin{array}{c}
h_x  \\
h_y  
\end{array} 
\right) \cdot
\left( h_x ^* ,  h_y ^* \right) dV \\
= -\frac{1}{2} \int_V   \omega
\left( \begin{array}{c}
\chi_{xx}^{''} h_x  +\chi_{xy}^{''} h_y  \\
\chi_{yx}^{''} h_x  +\chi_{yy}^{''} h_y 
\end{array} \right) \cdot
\left( h_x ^* ,  h_y ^* \right) dV \\
 \propto    \omega (\chi_{xx}^{''} h_x ^2 + \chi_{yy}^{''} h_y ^2),
\end{split}
\end{equation} 

The MW power absorption is thus given by the imaginary part of the diagonal elements $ \chi_{xx}^{''}$ and $\chi_{yy}^{''}$, for the field components in the x/y direction respectively. 

 Using that $\chi_{yy/xx}$ is written in the form $\chi_{yy/xx}=Z_1 / Z_2$, where $Z_i$ are complex numbers, one can seperate the real and imaginary part by multiplying the expression by the complex conjugate of the denominator: $\chi_{yy/xx} = \frac{Z_1 Z_2 ^*}{Z_2  Z_2  ^*}$. Assuming low damping, ($\alpha^2 \approx 0$) this gives:

\begin{equation}\label{eq:polder4}
\Re(\chi_{xx/yy})= \frac{ \omega_0 \omega_M(\omega_0^2 - \omega^2) - \beta_{x/y} \alpha \omega  \omega_M (\omega_0^2 + \omega^2)}{(\omega_0^2 - \omega^2)^2 + (2 \alpha \omega \omega_0)^2},
\end{equation}

\begin{equation}\label{eq:polder5}
\Im(\chi_{xx/yy})= \frac{-\alpha \omega  \omega_M (\omega_0^2 + \omega^2) - \beta_{x/y}  \omega_0 \omega_M(\omega_0^2 - \omega^2)}{(\omega_0^2 - \omega^2)^2 + (2 \alpha \omega \omega_0)^2}.
\end{equation}

As the FMR linewidth for permalloy films is small compared to the resonance frequency, one can assume that one does not need to deviate far from the resonance in order to observe the shape of the curve. That being the case, $\omega_0^2 + \omega^2 \approx 2\omega_0^2$, and

\begin{equation}\label{eq:fittingEq}
(\omega_0^2 - \omega^2)^2=(\omega_0+\omega)^2(\omega_0-\omega)^2\approx 4 \omega_0^2(\omega_0 - \omega)^2. 
\end{equation}

Hence, for narrow linewidths, Eq.(\ref{eq:polder5}) is well approximated by:

\begin{equation}\label{eq:polder6}
\Im(\chi_{xx/yy})\approx  \left( \frac{-\omega_M \Gamma_w }{4} \right) \frac{1 + \beta_{x/y}(\omega_0-\omega)/\Gamma_w}{(\omega_0-\omega)^2 + (\Gamma_w /2)^2},
\end{equation}

where the parameter $\Gamma_w = 2\alpha \omega$ has been introduced to describe the linewidth. This expression consists of two components: a symmetric absorption lineshape arising from the in-phase driving fields, and an antisymmetric dispersive lineshape proportional to $\beta$ arising from out-of-phase driving fields. The $\beta$ parameter is thus determined by the ratio between the absorptive and dispersive contributions to the FMR lineshape. \cite{ oates, poole}.

In our set-up the microwave frequency is fixed at 9.4 GHz, and the magnetic field $H_0$ is then swept to locate the ferromagnetic resonance at the resonance field, $H_0=H_R$, satisfying the condition for the resonance frequency, $\omega_R=\gamma \sqrt{H_R(H_R + 4 \pi M_s)}$. 
To extract $\beta$ from our experiments, we thus use an expression of same functional form as Eq. (\ref{eq:polder6}), but expressed in terms of field rather than frequency.

\begin{equation}\label{eq:fiteq}
\Im(\chi_{xx/yy}) = A \frac{1+ \beta_{x/y}(H_R-H_0)/\Gamma_\text{H}}{(H_R-H_0)^2 + (\Gamma_\text{H} /2)^2},
\end{equation}
\noindent
where A is an unimportant proportionality factor, and $\Gamma_\text{H}$ has been introduced to describe the linewidth.  
In this form, Eq.(\ref{eq:fiteq}) describes what is known in the litterature as Dysonian lineshapes \cite{dyson, oates, poole}.

In our experiments, we measure the field derivative of the MW absorption. The experimental data is thus fitted to the derivative of Eq. (\ref{eq:fiteq}) with respect to the external field, which is given by:

\begin{equation}\label{eq:fiteq2}
\begin{split}
&\frac{d }{dH_0} \Im( \chi_{xx/yy}) = A \Bigg [\frac{-\beta_{x/y}/ \Gamma_\text{H} }{(H_R-H_0)^2 + (\Gamma_\text{H} /2)^2}\\
 &+  \frac{2(H_R-H_0)[1+\beta_{x/y}(H_R-H_0)/ \Gamma_\text{H}]}{[(H_R-H_0)^2 + (\Gamma_\text{H} /2)^2]^2} \Bigg ].
\end{split}
\end{equation}

Through the $\beta$ parameter, FMR lineshapes in an otherwise unperturbed system is thus a measure of the amplitudes and relative phase of the MW field and the induced fields from eddy currents.

\section{Experimental setup}

Experiments were performed with Permalloy (Py=Fe$_{20}$Ni$_{80}$) as the ferromagnet layers, and gold (Au) and copper (Cu) as NM layers. The Py was grown by E-beam evaporation on oxidized silicon substrates, and the Au and Cu layers were grown by DC Magnetron sputter deposition. We controlled the thickness of the deposited NM layers using a Veeco Dektak 150 profilometer, and we cut our samples using a Dynatex DX-III combined scriber and breaker to obtain well defined sample geometries. Ferromagnetic resonance measurements were carried out in a commercial EPR setup (Bruker Bio-spin ELEXSYS 500, with a cylindrical TE-011 microwave cavity).

\begin{figure}[h]
\centering
\includegraphics[width=50 mm]{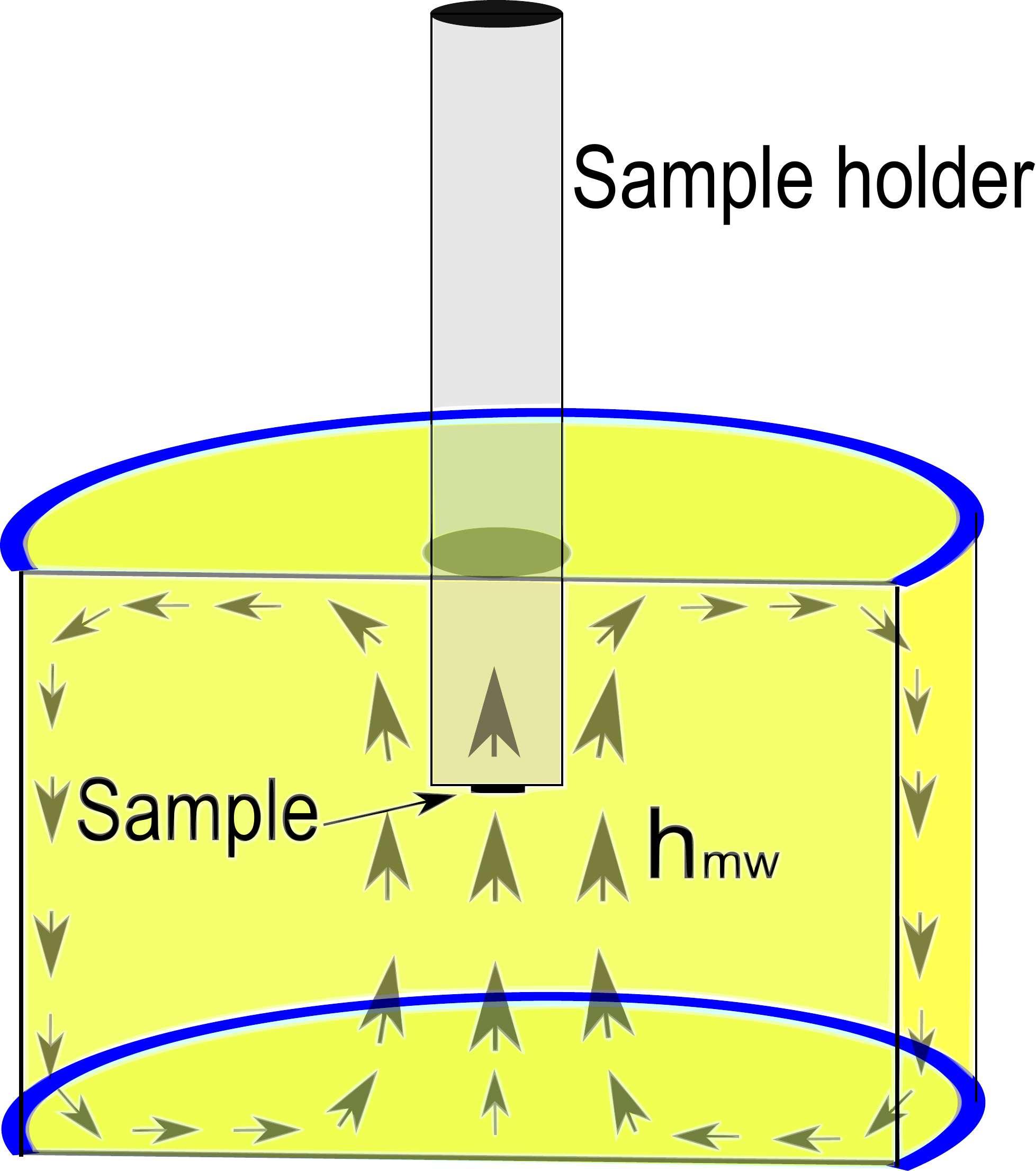}
\caption{\footnotesize Schematic of the cylindrical TE-011 microwave cavity, showing the sample position and field geometry.   }
\label{fig:cavity}
\end{figure}

The sample is attached to a quartz rod connected to a goniometer, allowing to rotate the sample 360 degrees. The MW field is oriented perpendicular to the sample plane  and is rotationally symmetric due to the cylindrical shape of the cavity, as shown in Fig. \ref{fig:cavity}.

Our FMR experiments were performed with a low amplitude ac modulation of the static field, which allows lock-in detection to be used in order to increase the signal to noise ratio. The measured FMR signal is then proportional to the field derivative of the imaginary part of the susceptibility. The experimental data was thus fitted to Eq.(\ref{eq:fiteq2}), $d\chi/ d H_0$
(i.e., we obtained an absorption line as in Fig.\ref{fig:lineshapes_sampgeom}b when the driving field had only the MW component; we obtained an absorption line as in Fig.\ref{fig:lineshapes_sampgeom}c when the driving field had a strong component from the eddy-current induced fields).

\section{Results and discussion}
\subsection{Effect of sample geometry}
\begin{figure}[h]
\centering
\includegraphics[width=70 mm]{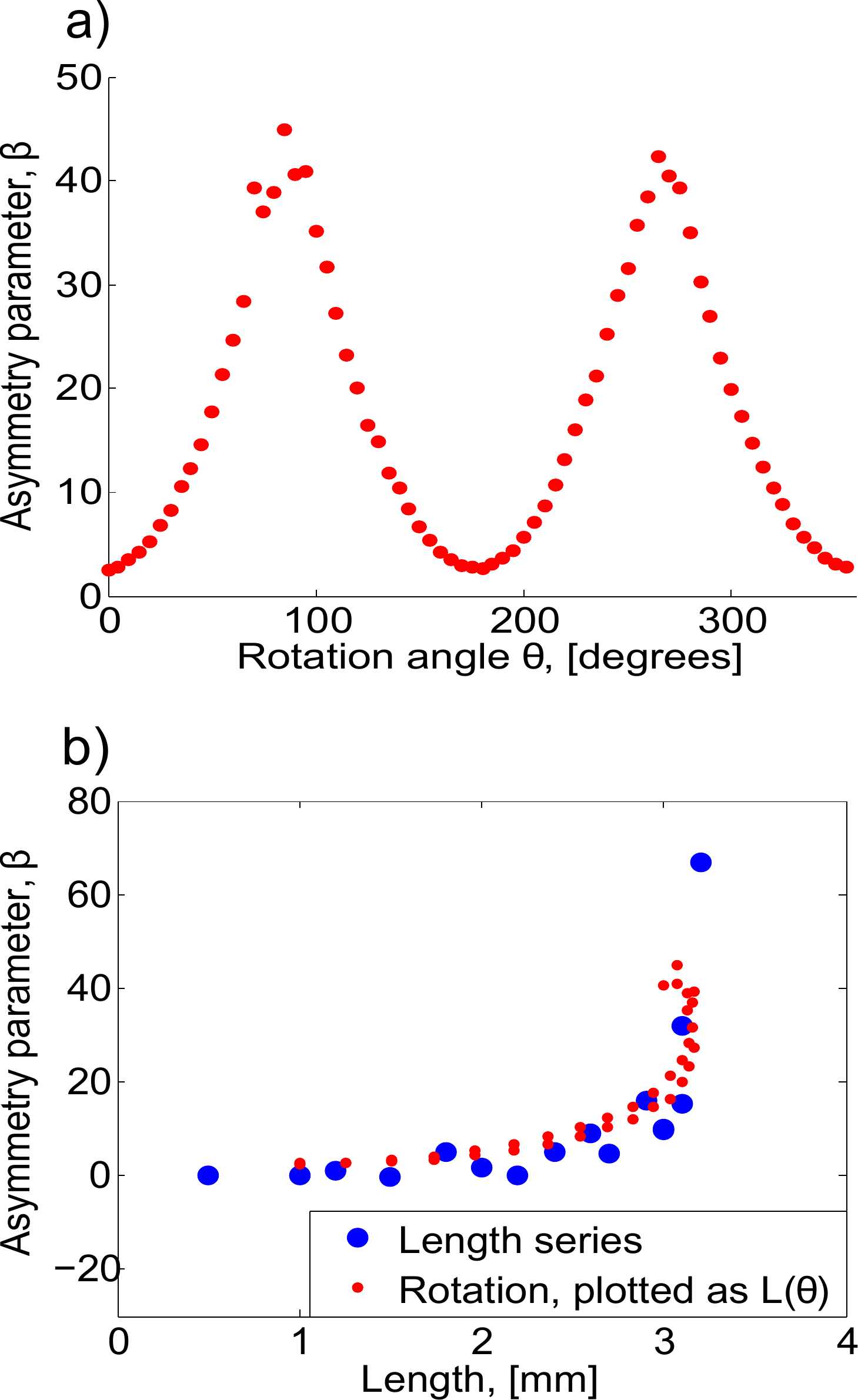}
\caption{\footnotesize  (a) Angular dependence of the $\beta$ parameter describing the FMR lineshape for a sample of 10 nm Py capped with 10 nm Au of dimension 1$\times$3 mm; the applied field is rotated 360 degrees in the film plane. (b) Sample length dependence of $\beta$ for samples  of 10 nm Py capped with 10 nm Cu (and 5nm Ta to prevent oxidation) of dimensions 1$\times L$ mm, i.e each datapoint in the "Length series" corresponds to a separate sample of length L. We also plotted in (b) the rotational measurements shown in (a) for a single sample, considering that the effective length in the direction of the applied field, $H_0$, is approximated by $L(\theta)=l\sin(\theta)+w\cos(\theta)$, $l=3$ mm and $w=1$ mm.}
\label{fig:rot_length_thick}
\end{figure}

We first focus on the effect of the sample geometry. A full in-plane 360 degrees rotation of a sample of dimensions 1$\times$3 mm with a thickness of 10 nm Py capped with 10 nm Au is shown in Fig.\ \ref{fig:rot_length_thick}a, where $\theta=0$ corresponds to an applied field, $H_0$, parallel to the short side of the sample.
We note that although capping the sample with a thin NM layer affects the lineshape asymmetry considerably, the resonance field $H_R$ and linewidth $\Gamma$ stay constant. Thicker NM layers of materials with considerable spin orbit coupling would lead to a linewidth broadening due to loss of spin angular momentum through spin pumping effects, but for thin Cu/Au layers this effect is negligible \cite{spinpump,spinpump2}.

The microwave field in the cavity can be considered uniform on the length scale of the sample, and rotationally symmetric due to its cylindrical shape. If the microwave excitaton is inhomogenous when rotating a long sample, it could be possible to excite magnetostatic modes in the FM film \cite{nonuniform}. However, if this was the case in our experiments, one should observe the same asymmetry for the FM without the NM capping layer. To rule out conclusively this as a cause of the asymmetry, we performed control experiments where we re-positioned the sample with an offset from the centre of the cavity (offset of the same order as the sample dimension). This did not affect the asymmetry of the lineshape, indicating that an inhomogenous MW field in the cavity could not be the cause of the observed effect.  

To investigate the effect of sample geometry further, a set of samples of dimensions 1$\times \text{L}$ mm, where L ranged from 0.5 to 4 mm was studied. The samples were again of 10 nm Py capped with 10 nm Cu, and 5 nm Ta to prevent oxidation. We notice that the sample length in the direction parallel to the applied field is the main parameter that determines the asymmetry of the FMR lineshapes, given by the parameter $\beta$. Figure \ref{fig:rot_length_thick}b shows the dependence of sample length when the varying dimension is parallel to the applied field. We see that the asymmetry increases with the sample's length and reaches a value where $\beta$ appears to diverge at a length of about 3.3 mm. Samples with a length below 1 mm have lineshapes almost identical to samples with no NM capping.

We consider now the basic physics to describe the above results. The induced eddy currents flow in closed loops in planes perpendicular to the MW magnetic field, which is perpendicular to the film plane in our experiment. Thus, to obtain circulating eddy currents as shown in  Fig. \ref{fig:lineshapes_sampgeom}a, it is required to have the MW field perpendicular to the film plane.  We have conducted control experiments where the MW fields were applied in the film plane and we observed that the FMR lineshapes were always symmetric, indicating there were no observable effect of the eddy currents. 

In our experimental geometry, the induced eddy currents flow mainly in circulating paths, with highest current density along the sample edges \cite{ind_eddycurr, ind_eddycurr2}. The induced Oe fields have a component in the film plane and another perpendicular to the film plane. As indicated in Fig. \ref{fig:lineshapes_sampgeom}a, for the sample edges that are parallel to the applied field, the Oe fields will have the main in-plane component perpendicular to the applied field and could thus affect the FMR of the Py film. On the other hand, currents along sample edges perpendicular to the applied field will give rise to an in-plane Oersted field that is parallel to the applied field, and should not affect the FMR response. 

The observed strong rotational dependence (See, Fig.\ \ref{fig:rot_length_thick}a), suggest that the effective driving field has the dominating contribution oriented in the sample plane; the contribution from the component perpendicular to the sample plane should not depend on the direction of the sample edges with respect to the applied field. As the effective driving field appears to be dominated by the in plane components, this indicates that the induced local field perturbing the FM is larger than the external field. This could be possible due to the close proximity to the induced currents at the FM/NM interface.

We now compare the length series with the rotational measurements by using a simple geometric approximation: we consider that the length of the sample parallel to the applied field is given by $L(\theta)=l\sin(\theta)+w\cos(\theta)$, where $l$ and $w$ are the length (3 mm) and width (1 mm) of the sample, and $\theta=0$ corresponds to the applied field parallel to the short side of the sample. We have plotted in Fig. \ref{fig:rot_length_thick}b the rotational measurements following this approach and we can see that the resulting curve is almost identical to the length series.

\begin{figure}[ht]
\centering
\includegraphics[width=80mm]{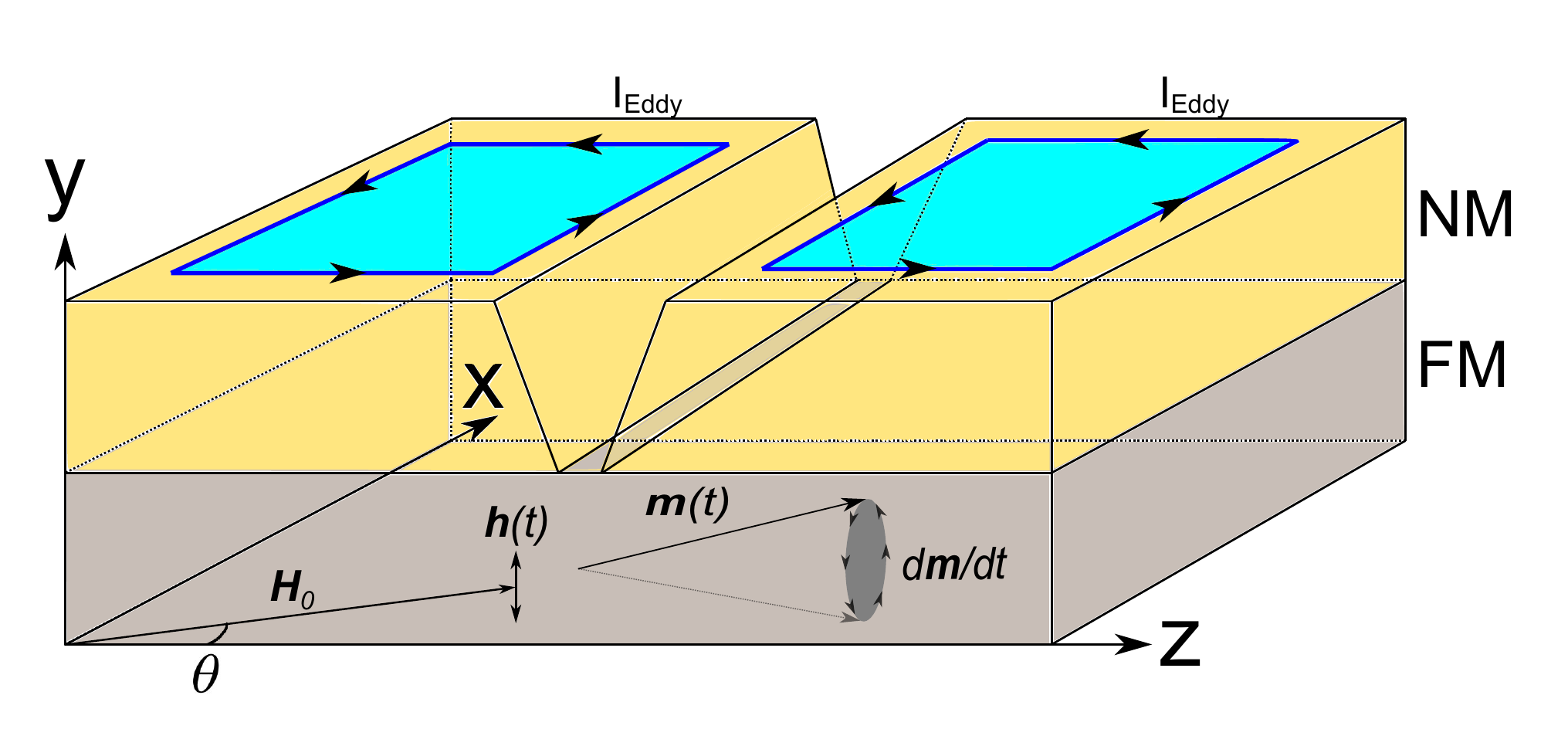}
\caption{\footnotesize Scratching the sample limits the size of the posible current loops, reducing the magnitude of the induced fields. }
\label{fig:sample_geom_split}
\end{figure}

To investigate the effect of sample size closer, we designed a control experiment that consisted of taking a large sample of dimensions 1$\times$3 mm and dividing it into electrically isolated regions of 1$\times$1 mm. This was performed using an automated scriber that scratched the sample without breaking it---we limited the size of the possible current loops (as illustrated in Fig. \ref{fig:sample_geom_split}). 

We tested this for samples with no NM capping, and the FMR signal was not affected. However, the same procedure on a sample capped with 10 nm Au presented a remarkable effect: the lineshape before scratching the sample was strongly asymmetric, but after scratching the film it returned to being symmetric again and matched the lineshapes for a sample of dimensions 1$\times1$ mm.

The asymptotic behaviour of $\beta$ as the sample length increases, can be understood by considering how the sample size affects the magnitude of the induced field. As a simplified model, we approximate the current path as a rectangular loop around a sample of length $l$ and width $w$. The induced electromotive force (EMF) is then given by the rate of change of magnetic flux through the area enclosed by the loop; its absolute value is given by $|\epsilon|=lw\displaystyle \left|\frac{\partial h}{\partial t}\right| \propto lw~h_{\text{MW}} 2\pi f$, where $f$ and $h_{MW}$ are the MW frequency and amplitude respectively. The resistance of such a loop is given by $R=2 R_s (l+w)/\zeta$, where $R_s$ is the sheet resistance and  $\zeta$ is the width of the current path. The induced current is finally given by $I=\epsilon/R$. We consider as an approximation that the magnetic field resulting from a current $I$ in such a plane is given by $h_{\text{ind}}=\mu_0 I /2\zeta$, where $\mu_0$ is the vacuum permeability. The expression for the induced field is thus proportional to the sample size.

As an estimate for the strength of the induced field, we calculated this for a square sample, $w=l$, using the above expression for $h_\text{ind}$. 
 The sheet resistance was measured to be $R_s \approx 50 \ \Omega$ for samples with 10 nm Py capped by 10 nm Cu and 5 nm Ta. Considering the width of the current path along the sample edge as $\zeta=w/4$, one can estimate the sample size where $h_\text{ind}=h_\text{MW}$.
In the presence of a MW field of 9.4 GHz and 6 $\mu$T (values used during our experiments), we obtained that at a dimension of about 2$\times$2 mm the induced field equals the MW field. Our estimate corresponds with what we see in the experiments: when the sample size approaches the mm scale, the effects become increasingly important.

\subsection{Thickness of NM layer}

\begin{figure}[ht]
\centering
\includegraphics[width=80mm]{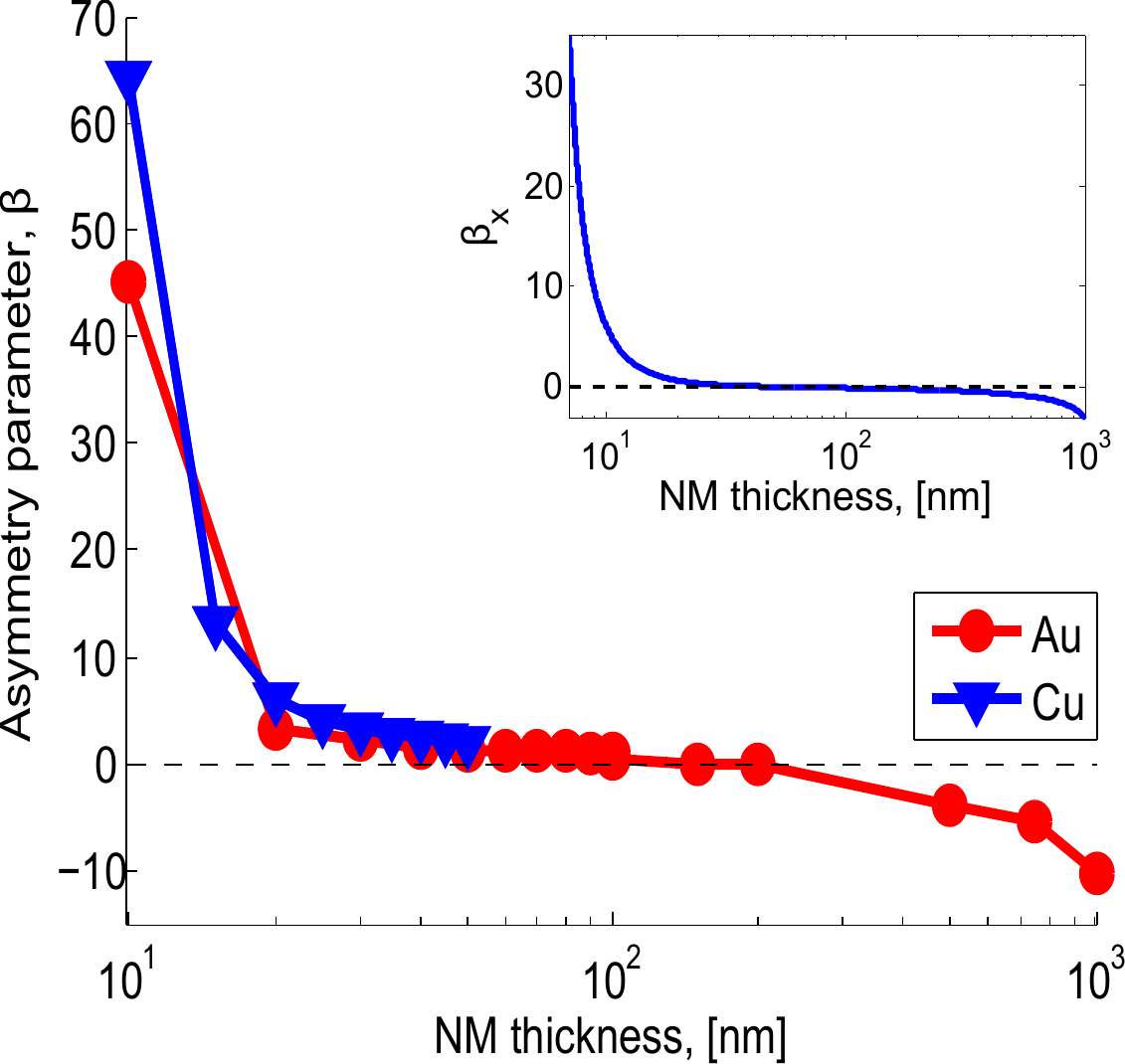}
\caption{\footnotesize NM thickness dependence of the $\beta$ parameter describing the FMR lineshape for a sample of dimension 1$\times$3 mm, with the applied field parallel to the long side of the sample. Comparing Au and Cu as NM layer in the region 0-50 nm. Inset: Calculated thickness dependence of $\beta_x$, given by Eq. (\ref{eq:beta2}) and  (\ref{eq:phase})}
\label{fig:thickness_phase}
\end{figure}

Next, we focus on another important parameter that governs the effect of eddy currents: the thickness of the NM layer. We prepared samples with a NM (both with Au and Cu) thickness ranging from 10 nm to 1 $\mu$m (Au) and 10-50 nm (Cu). (The experiments presented here were also performed in two more samples, where we obtained the same results. The measured asymmetry parameter $\beta$ as a function of NM thickness is shown in Fig. \ref{fig:thickness_phase} for a sample of dimensions 1$\times3$ mm, with the applied field parallel to the long side of the sample. Replacing the Au layer by Cu in the range 10-50 nm shows a similar behavior; the thicker the NM, the more symmetric the FMR lineshapes. In the thick film limit one observes asymmetric lineshapes again, but with an opposite sign of the $\beta$ parameter.

To explain the thickness dependence, we use a simplified model where we assume that the induced eddy currents circulate in two dimensional planes in the NM layer. The Oersted fields originated by the eddy currents have a relative phase lag, $\phi$, compared to the external MW field, which in the ideal case of no inductance is expected to be $\phi=-90$ degrees ($I_\text{Eddy} \propto \frac{\partial h}{\partial t}$). However, due to the inductance and resistance of the NM film, there will be an additional phase between the MW field and the induced field that depends strongly on the NM thickness (due to the low conductivity of Py compared to Au/Cu, we consider the currents to circulate mainly in the NM layer). At larger thicknesses, one also needs to take into account phase shifts due to the skin effect. Considering this, one can write the relative phase lag as a function of NM thickness as \cite{jackson}:  

\begin{equation}\label{eq:phase}
\phi(d)= -\left[90  + \tan^{-1} \left( \frac{\omega L(d)}{R(d)} \right) + d/\delta \right],
\end{equation}
where $\omega$ is the microwave angular frequency, $L$ and $R$ are the inductance and resistance of the film, $d$ is the NM thickness and $\delta$ is the MW skin depth ($\simeq 800$ nm for Au at 10 GHz). 

To estimate values for the inductance, we consider a rectangular current path along the edges of the NM layer \cite{ inductance}, and sample dimensions of $w$=1 mm, $l$=3 mm with thickness $d$  ($L \approx 10^{-8}$ H  for a thickness of 10 nm). 

\begin{equation}\label{eq:induction}
\begin{split}
L(d)= \frac{\mu_0 \mu_r}{\pi} \Biggl[ -2(w+l) +2 \sqrt{l^2+w^2}-l \cdot ln \left( \frac{l + \sqrt{l^2+w^2}}{w} \right) \\  
- w \cdot ln \left( \frac{w + \sqrt{l^2+w^2}}{l} \right) + l\cdot ln \left( \frac{2l}{ \zeta d}\right) + w \cdot ln  \left( \frac{2w}{ \zeta d}\right)  \Biggr],
\end{split}
\end{equation} 
\noindent
where $\mu_R$ is the relative permeability of the NM film ($\approx 1$), and $\zeta$ is the width of the current path, set to $w/2$ in this calculation. 

As mentioned in the previous section, we measured a sheet resistance of $R_s \approx 50 \ \Omega$ for samples with 10 nm Py capped by 10 nm Cu and 5 nm Ta. However, for NM layers in this thickness regime, the conductivity depends strongly on the thickness. This is due to increased interface scattering in thin films when the thickness is of the same order as the electron mean free path. Due to this we estimate the film resistance, $R(d)$, as a function of thickness by introducing a correction factor, $\eta$, which describes a correction to the film conductivity compared to its bulk value. In \cite{screening2} it was found that the increase in conductivity is close to linear in film thickness below 20-30 nm, before reaching an asymptotic value for thicker films. In our calculations we thus considered $\eta$ to be a linear function of the film thickness below 20nm: $\eta=\text{min}\left(1,\frac{d}{l_{\text{mfp}}}\right)$, where $d$ is the NM thickness and we set $l_{\text{mfp}}=20$ nm.

From the rotational measurements, we argued that the effective driving field is dominated by the in plane component of the induced field. For thicker NM layers the sheet resistance is reduced, and the magnitude of the induced field should thus increase, as $|h_\text{ind}| \propto 1/R_s$. Due to this, the effective driving field should be dominated by the in plane component, $h_x$, also for thicker NM layers. 
From Eq. (\ref{eq:Pabs2}) and (\ref{eq:fiteq2}), the asymmetry of the lineshape is then given by the parameter $\beta_x$.

Using these approximations we computed the phase shift between the MW field and the induced field, and calculated the thickness dependence of $\beta_x$, given by Eq. (\ref{eq:beta2}) and  (\ref{eq:phase}),  illustrated in the inset of Fig. \ref{fig:thickness_phase}.
As the NM thickness increase, the phase difference approaches a value of 180 degrees, which then corresponds to $\beta_x=0$ (i.e., the asymmetric lineshapes disappear quickly as the NM thickness increases). For thicker NM layers, one gets an additional contribution to the phase difference due to the skin effect. At a certain thickness the phase shift will thus be larger than 180 degrees, which corresponds to an opposite sign of $\beta_x$. 

These main features of our simple model agrees well with the experimental data in Fig. \ref{fig:thickness_phase}, where the asymmetry drops off quickly with the thickness for thin NM layers. As the thickness of the NM layer is increased further, one also observes the expected transition to asymmetric lineshapes again, but with an opposite sign of the $\beta$ parameter. The thick film limit corresponds to the regime where one usually assumes eddy-current effects to become important, i.e. when the NM thickness approach its MW skin depth.  

Experimentally, we observed the strongest lineshape asymmetry in films with a NM thickness of 10 nm. We also investigated thinner NM layers of 5 nm, and the FMR lineshapes were similar to single Py films. We believe this is because we had non-continuous metal films for these thicknesses; Au films tend to be granular and the Cu films might have oxidized. 

\section{Summary}
To summarize, we have shown that induced eddy currents can play an important role in FM/NM bilayer structures for certain sample geometries. In contrast to what is usually assumed about eddy currents, our results indicate that these effects can be important also for film thicknesses far below their skin depth.
In FMR measurements, the influence on lineshape asymmetries has to be taken into account for NM layers below 50 nm and sample dimensions above approx. $1$ mm$ ^2$ when the MW field has a significant component perpendicular to the film plane. 

The dynamics of the system is determined by the interplay of the MW fields and induced fields by eddy currents, and we have shown that this coupling is tunable through changing the sample geometry and the NM layer thickness. The tunability of the coupling opens up possibilities to use patterned NM structures to tailor the local field geometry and phase of the induced microwave fields, which could be of importance for magnonics applications. 
\\

\section*{Acknowledgements}
We are grateful for insightful discussions with A. Brataas and H. Skarsv\r{a}g.
This work was supported by the Norwegian Research Council (NFR), project number 216700. 
V.F acknowledge long term support from NorFab Norway, and partial funding obtained from the Norwegian PhD Network on Nanotechnology for Microsystems, which is sponsored by the Research Council of Norway, Division for Science, under contract no. 221860/F40. FM acknowledges support from Catalan Government through COFUND-FP7 and support from MAT2011-23698. A. D. K acknowledges support through the US-National Science Foundation, NSF-DMR-1309202.


\begin{thebibliography}{21}

\bibitem{eddy1}
C. Kittel,
Physical review. 73, 2, (1948)

\bibitem{eddy2}
P. Pincus,
Physical review. 118, 3, (1960)

\bibitem{spindyn1}
Y. Tserkovnyak, A. Brataas, G. E. W. Bauer, B. I. Halperin,
Rev. Mod. Phys. 77, 1375, (2005)


\bibitem{spindyn2}
V. V. Kruglak, S. O. Demokritov, D. Grundler,
J. Phys. D: Appl. Phys 43 264001, (2010).  


\bibitem{screening}
S. Fahy, C. Kittel, S. G. Louie, 
American Journal of Physics 56, 989 (1988):

\bibitem{screening2}
I. V. Antonets, L. N. Kotov, S. V. Nekipelov, and E. N. Karpushov, 
Technical Physics, 
November 2004, Volume 49, Issue 11, pp 1496-1500

\bibitem{screening_FMR}
M. Bailleul, 
Appl. Phys. Lett. 103, 192405 (2013)

\bibitem{screening_FMR2}
M. Kostylev,
J. Appl. Phys. 106, 043903 (2009)

\bibitem{screening_FMR3}
I. S. Maksymov and  M. Kostylev,
J. Appl Phys. 116, 173905 (2014)

\bibitem{screening_FMR4}
I. S. Maksymov, Z. Zhang, C. Change, and M. Kostylev,
IEEE MAGNETICS LETTERS, Volume 5 (2014)


\bibitem{review_FMR}
M. Harder, Z. X. Cao, Y. S. Gui, X. L. Fan, and C.-M. Hu, 
Phys. Rev. B 84, 054423 (2011)


\bibitem{spinHall}
O. Mosendz, V. Vlaminck, J. E. Pearson, F. Y. Fradin, G. E. W. Bauer, S. D. Bader, and A. Hoffmann, 
Phys. Rev. B 82, 214403 (2010)

\bibitem{spinHall2}
O. Mosendz, J. E. Pearson, F. Y. Fradin, G. E. W. Bauer, S. D. Bader, and A. Hoffmann, 
Phys. Rev. Lett. 104, 046601 (2010)

\bibitem{spinHall3}
L. Liu, T. Moriyama, D. C. Ralph, and R. A. Buhrman, 
Phys. Rev. Lett. 106, 036601 (2011)

\bibitem{spinHall4}
S. I. Kim, M. S. Seo, S. Y. Park
J. Appl. Phys. 115, 17C501, (2014)



\bibitem{ind_eddycurr}
M.Krakowski, 
Archiv f\"{u}r Elektroteknik 64 (1982) 307-311

\bibitem{ind_eddycurr2}
G.De Mey, 
Archiv f\"{u}r Elektrotechnik,
1974, Volume 56, Issue 3, pp 137-140

\bibitem{LLG}
L. Landau, E. Lifshitz.
Phys. Z. Sowjetunion 8, 153 (1935).

\bibitem{polder}
D. Polder
Phil. Mag. 40, 99-115 (1949)

\bibitem{oates}
C. J. Oates, F. Y. Ogrin, S. L. Lee, P. C. Riedi, G. M. Smith, T. Thomson
J. Appl. Phys. 91, 1417, (2002).

\bibitem{poole}
C. P. Poole, "Electron Spin Resonance-A Comprehensive Treatise on Experimental techniques".
Wiley, New York, (1967).


\bibitem{dyson}
F. J. Dyson
Phys. Rev. 98, 349 (1955)

\bibitem{spinpump}
Y. Tserkovnyak, A. Brataas and G. W. Bauer,
Phys. Rev. B. 66, 224403 (2002)

\bibitem{spinpump2}
S. Mizukami, Y.Ando and T. Miyazak,
Phys. Rev. B. 66, 104413 (2002)



\bibitem{nonuniform}
P. Wolf, Z. Angew,
Phys. 14, 212 (1962)


\bibitem{jackson}
J. D. Jackson, 
Classical Electrodynamics, (1998)


\bibitem{inductance}
Frederick W. Grover,
Inductance Calculations,
(2004)

\end{thebibliography}
\end{document}